\documentclass{aastex}

\newcommand{\kms}   {km~s$^{-1}$}
\newcommand{\mjy}   {mJy~beam$^{-1}$}
\newcommand{\Mjy}   {$\mu$Jy~beam$^{-1}$}

\newcommand{\lo}    {$L_{\sun}$}
\newcommand{\mo}    {$M_{\sun}$}

\newcommand{\et}    {et al.}
\newcommand{\ie}    {i.\,e.,}
\newcommand{\eg}    {e.\,g.,}

\newcommand{\ro}    {Rodr\'{\i}guez}
\newcommand{\xx}    {$\times$}
\newcommand{\mm}    {$\pm$}
\newcommand{\D}[2]  {\mbox{#1$\arcdeg \pm$#2$\arcdeg$}}
\newcommand{\DS}[3] {\mbox{#1$\farcs$#2(#3)}}

\slugcomment{Submitted to The Astronomical Journal}
\shortauthors{Girart et al.}
\shorttitle{The YLW 15 Binary System}

\begin{document}

\title{On the Evolutionary State of the Components of the YLW 15 Binary System}

\author{
J. M. Girart\altaffilmark{1,2}, 
S. Curiel\altaffilmark{3}, 
L. F. Rodr\'{\i}guez\altaffilmark{4}, 
M. Honda\altaffilmark{5}, 
J. Cant\'o\altaffilmark{3}, 
Y. K. Okamoto\altaffilmark{6}, and 
S. Sako\altaffilmark{5}}

\altaffiltext{1}{
Departament d'Astronomia i Meteorologia, Universitat de Barcelona, Av.
Diagonal 647, 08028 Barcelona, Catalunya, Spain
}
\altaffiltext{2}{
Institut de Ci\`encies de l'Espai (CSIC)-IEEC, Gran Capit\`a 2,
08034 Barcelona, Catalunya, Spain;
{\tt{girart@ieec.fcr.es}}
}
\altaffiltext{3}{
Instituto de Astronom\'{\i}a, Universidad Nacional Aut\'onoma de
M\'exico, Apdo. Postal 70-264, 04510 M\'exico D.F., M\'exico
{\tt{scuriel@astroscu.unam.mx}}
}
\altaffiltext{4}{
Centro de Radioastronom{\'\i}a y Astrof{\'\i}sia, UNAM, 
Apdo. Postal 3-72, (Xangari), 58089 Morelia, Michoac\'an, M\'exico
{\tt{l.rodriguez@astrosmo.unam.mx}}
}
\altaffiltext{5}{Subaru Telescope, National Astronomical Observatory of Japan, 
650 North A'ohoku Place, Hilo, HI 96720
{\tt{hondamt@subaru.naoj.org}}
}
\altaffiltext{6}{
Institute of Physics, Center for Natural Science, Kitasato University 1-15-1
Kitasato, Sagamihara, Kanagawa 229-8555, Japan}


\begin{abstract}
We report centimeter continuum observations with the VLA and the VLBA as well
as mid-infrared observations with COMICS/SUBARU toward the components of the 
YLW~15 very young binary system, VLA~1 and VLA~2.  The centimeter emission of
the two components traces partially thick free-free emission, likely due to
collimated, ionized winds. VLA~1 is an embedded protostar, undetected in the
near-IR, and possibly in the Class 0 to Class I transition and powering a
Herbig-Haro outflow. Its mid-IR emission appears slightly resolved with a
diameter of $\sim 16$~AU, 
possibly tracing circumstellar material from both the envelope and the disk. 
VLA~2 is a typical Class I object, unresolved in the mid-IR, and is the
responsible of the strong X-ray emission associated with YLW~15. The expected
centimeter ''peri-stellar'' emission associated with the X-ray emission is not
detected with the VLBA at 6~cm likely due to the high optical depth of the
free-free emission.  Strikingly, the near to mid-IR properties of YLW~15
suggest that VLA~1 is a more embedded YSO, or alternatively, less luminous 
than VLA 2, whereas orbital proper motions of this binary system indicate that
VLA 1 is more massive than VLA 2. This result is apparently against the
expected evolutionary scenario, where one expects that the more massive YSO in
a binary system is the more evolved and luminous YSO. Finally, the nearby source 
YLW~16A is detected with the VLA, its position coincides with reported near-IR and 
X-ray sources. 
\end{abstract}

\section{Introduction\label{intro}}

The $\rho$ Ophiuchi molecular cloud complex is one of the nearest star
formation regions ($\sim$ 120~pc: Knude \& Hog 1998) and is also one of the
best studied low mass star forming regions.  YLW~15 (IRS 43) is a young stellar
object (YSO) embedded in the Oph-F dense core, located in the southeastern
region of the molecular cloud complex (\eg\ Motte, Andr\'e \& Neri 1998).  The
near-IR to mid-IR (\eg\ Wilking, Lada, \& Young 1989), far-IR (\eg\ Bontemps
\et\ 2001; Pezzuto \et\ 2002;  Nisini, Gianni \& Lorenzetti 2002) and millimeter 
(\eg\ Andr\'e \& Montmerle 1994; Bontemps \et\ 1996)
properties suggest that YLW~15 belongs to the Class I stage of evolution.  The
spectroscopic features in the near-IR show that YLW~15 is a YSO  with K5
spectral type and with dwarf like surface gravity, \ie\ a luminosity class
IV/V, having a stellar mass of 0.5~\mo\ (Greene \& Lada 2002).   It has been
proposed that YLW~15 constitutes a wide-binary system with GY~263, located 
6$''$ (720~AU) northwest of YLW~15 (\eg\ Allen \et\ 2002).  YLW~15 powers a 
compact molecular outflow (Bontemps \et\ 1996) and a Herbig-Haro outflow 
(\eg\ Grosso \et\ 2001).

YLW~15 is one of the youngest low mass stars associated with X-ray emission. 
First detected by ROSAT (Casanova \et\ 1995), the X-ray emission shows
quasi-periodic flares (Tsuboi \et\ 2000).  A ``super-flare''  was observed
during a few hours (Grosso \et\ 1997).   Grosso \et\ (1997) suggested that the
flare arises from a magnetically confined, low-density plasma bubble with a
diameter of $~\sim0.05$--0.3~AU around the young star.   The quasi-periodic
X-ray flares of YLW~15 were explained by Montmerle \et\ (2000) as due to ``fast
rotation of the central star with respect to the accretion disk, which results
in star-disk shearing of the magnetic lines, producing magnetic reconnection
and mass loss, and eventually extremely high X-ray luminosities'', which is
different from the ``classical'' quasi steady-state situation expected for the
T Tauri stars.  

Its radio continuum emission has been well studied at a moderate angular
resolution ($\sim 10''$) showing no variability (\eg\ Leous \et\ 1991).  VLBI
milliarcsecond angular resolution observations  of YLW~15 show that the
emission is resolved out at ``peristellar'' scales  (\ie\  its size is larger
than $1.2\times10^{13}$~cm), suggesting that the  emission is thermal in origin
and probably comes from circumstellar ionized  winds (Andr\'e \et\ 1992).
Sub-arcsecond angular resolution VLA observations revealed that the radio
emission arises from a binary system, YLW~15~VLA~1 and YLW~15~VLA~2 (Girart,
\ro\ \& Curiel 2000, Paper I).  In these maps VLA~1 was clearly elongated with
a position angle of PA$=25\arcdeg \pm 2\arcdeg$.  This elongation aligns well
with HH~224NW1 and a chain of near-IR HH knots (Grosso \et\ 2001).  VLA~2
appeared unresolved down to scales of $\sim 0\farcs4$.  In Paper I we
suggested  that the two radio sources trace a binary young stellar system with
each component possibly having a different evolutionary status.  Using a 12
year time baseline of subarcsecond angular resolution VLA observations, Curiel
\et\ (2003, Paper II) have measured absolute proper motions of 24~mas~yr$^{-1}$
in YLW~15.  In addition, the VLA~1 and VLA~2 relative astrometry analysis done
in Paper II reveals orbital proper motions that suggest a total mass of the
binary system of $\ga$1.7~\mo.  These results show also that VLA~1 is likely
more massive than VLA~2.

In this paper we present VLA sub-arcsecond angular resolution observations at
3.6 and 6~cm toward YLW~15, as well as VLBA 6~cm and 
COMICS/SUBARU 8.8~$\mu$m observations.  In \S~2 and \S~3 we give the 
details of the observations and results with the VLA, VLBA and SUBARU. In 
\S~4 we discuss the nature of the radio emission and compare its properties with 
those obtained at other wavelengths in order to elucidate the nature of this interesting 
binary protostellar system.

\section{Observations}

\subsection{VLA}

The subarcsecond radio continuum observations were carried out with the Very
Large Array (VLA) of the National Radio Astronomy  Observatory\footnote{NRAO is
a facility of the National Science Foundation  operated under cooperative
agreement by Associated Universities, Inc.}\ between 2000 and 2002: 2000
December 27 (3.6 and 6~cm), 2002 February 10 (3.6~cm), 2002 March 5 (3.6~cm)
and 2002 March 8 and 9 (6~cm).  The absolute amplitude and phase calibrator
were always 3C286 and 1626$-$298, respectively.   The phase center for all the
observations was $\alpha$(J2000)$=16^{\rm h}27^{\rm m}26\fs920$,
$\delta$(J2000)$=-24\arcdeg40'50\farcs40$.  The data were edited and calibrated
following the standard VLA procedures with the AIPS software package of NRAO.  
Maps were done using robust weighting of 0.0 and $-$0.5 for the 3.6 and 6~cm
observations, respectively. The 3.6~cm flux measured for VLA~1 and VLA~2 were
the same, within the uncertainties, in the 2002 February 10 and March 5
epochs.  The same happened for the 6~cm fluxes in the 2002 March 8 and 9
epochs. Therefore, and given the proximity of the different observations in 
2002, the February 10 and March 5 visibility data sets (3.6~cm) were combined
in order to improve the sensitivity of the final maps.  The same was done for
the March 8 and 9 visibility data sets (6~cm).   All the fluxes were corrected
by the VLA antenna primary beam response.  Table~\ref{tobsA} shows the achieved
synthesized beam and the rms noise achieved for the 2000 and 2002 epochs.

\subsection{VLBA}

The milliarcsecond radio continuum observations at 6 cm (4.96 GHz) were carried
out using the 10 antennas of the Very Long Baseline Array (VLBA) of the
National Radio Astronomy  Observatory on two epochs, 2001 December 14 and 2002
March 22.  The observations were done in the phase-referencing mode using one
polarization.  The phase center of the observations was 
$\alpha$(J2000)$=16^{\rm h}27^{\rm m}26\fs9400$, 
$\delta$(J2000)$=-24\arcdeg40'50\farcs773$. Left-circular polarization was
recorded using 2 bit sampling across bandwidths of 8 MHz at each frequency. The
VLBA correlator produced 16 frequency channels across each 8 MHz bandwidth. 
Eight IFs were used, which give a total bandwidth of 64 MHz.  Fringe fitting
was performed observing the calibrator J1625$-$2527, located within a degree
from YLW~15.  The calibration and reduction of the data was performed following
the standard VLBA procedures with AIPS (Ulvestad 1991).  Maps were done using
IMAGR with natural weighting in order to maximize the sensitivity. 
Table~\ref{tobsB} shows the rms noise and the synthesized beam for each epoch.

\subsection{COMICS/SUBARU}

The mid-infrared imaging observations of YLW~15 at 8.8~$\mu$m were carried out
with the COMICS (COoled Mid-Infrared Camera and Spectrometer; Kataza et al.
2000) on the SUBARU Telescope on 2001 June 29 at a chopping frequency of
0.45~Hz. HD~14~5897 was also observed as a flux standard star. Total on-source
integration time was 33.9 sec for the YLW~15 image. After standard reduction
procedures, we employed the shift-and-add method to achieve the
diffraction-limited image. The FWHM for the PSF star was $0\farcs28$. Aperture
photometry was performed using the PHOT routine within IRAF. For the HD~14~5897
(the flux standard star) photometry an aperture of 6 pixels ($0\farcs78$) was
used.  The flux adopted for HD~14~5897 at 8.8~$\mu$m is 7.43~Jy. For the YLW~15
photometry, an aperture radius of 4 pixels ($0\farcs52$) was used.  In order to
estimate the total flux of YLW~15 VLA~1 and VLA~2, we first did the aperture
photometry for VLA~2, which has a peak intensity about 5 times stronger than
VLA~1. The flux measured for VLA~2 is 1.49~Jy at 8.8~$\mu$m. For VLA~1, we
first suppressed the VLA~2 emission: this was done by rescaling and shifting 
the HD~14~5897 image to match it with the VLA~2 intensity and position. 
Second, the resulting image was subtracted from the YLW~15 image. The 
aperture photometry on VLA~1 was done on the YLW~15 image where VLA2 was 
suppressed. The flux measured for VLA~1 is 0.356~Jy.

\section{Results}

\subsection{YLW~15}

\subsubsection{VLA\label{vla}}

\begin{figure}
\epsscale{0.8}
\plotone{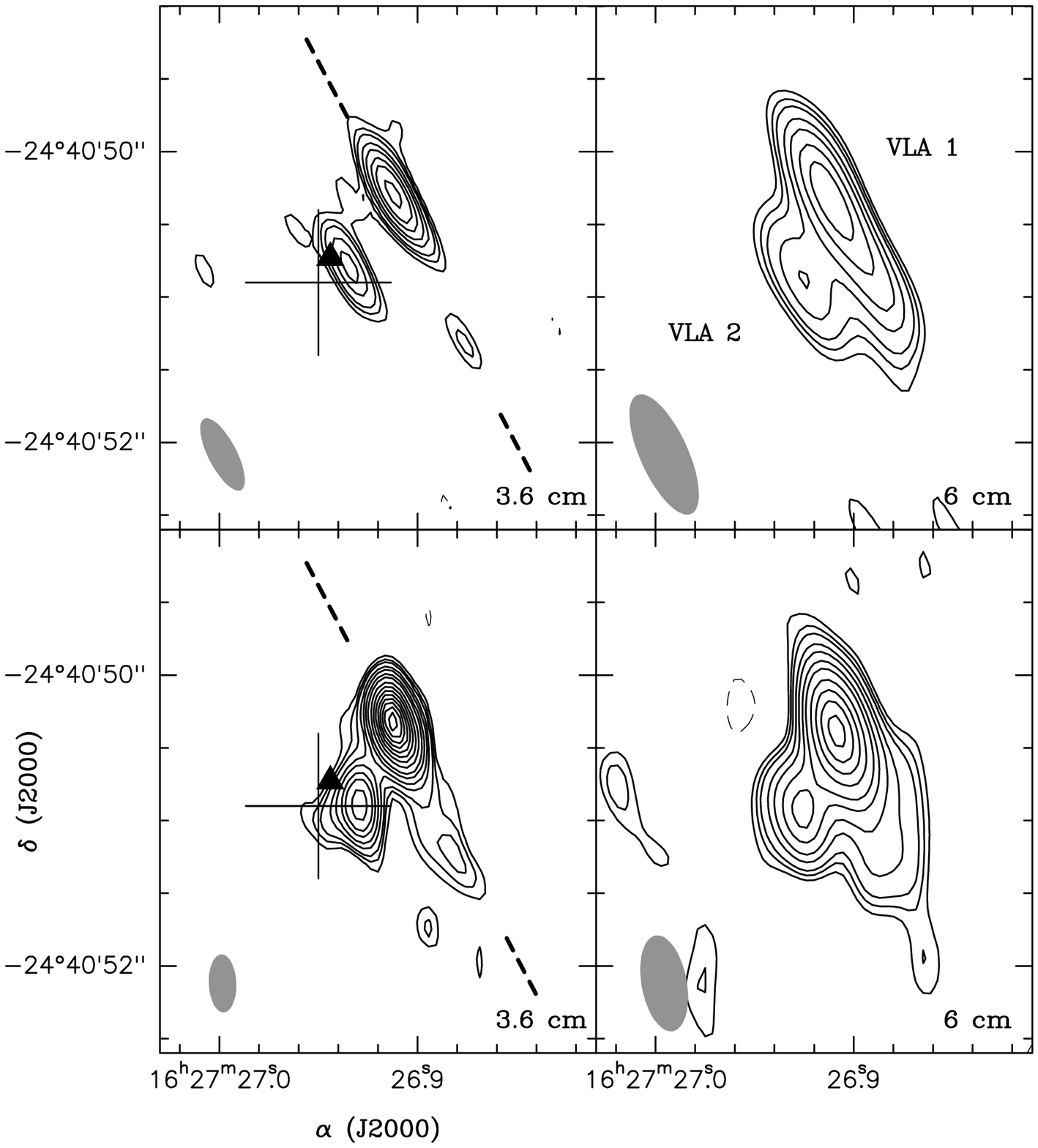}
\caption
{VLA continuum images of YLW 15 at 3.6 cm (left panels) and 6 cm (right
panels), for the epochs 2000 (top panels) and 2002 (bottom panels). Contours
are  $-$3, 3, 4, 5, 7 9, 12, 15, 20,\nodata, 55$\times$ the rms noise of the
images  (50~\Mjy\ at 3.6 cm and 59~\Mjy\ at 6~cm for the 2000 epoch, 17~\Mjy\
at 3.6 cm and 26~\Mjy\ at 6~cm for the 2002 epoch observations). The half power
size of the synthesized beams are shown in the bottom left corner (see
Table~\ref{tflux} for their sizes). The cross on the left panels shows the
position uncertainty (Imanishi private communication) of the X-ray source no.
54 (Imanishi et al.  2001).  The filled triangle shows the position of the
near-IR  2MASS~1627269-244050 source.  The dashed line shows the direction to
the HH outflow associated apparently with VLA~1 (Grosso et al. 2001).
}
\label{fmapA}
\end{figure}

Figure~\ref{fmapA} shows the 3.6 and 6~cm subarcsecond angular resolution maps
of the two epochs, 2000 and 2002. Table~\ref{tflux} gives the positions, flux
densities and the deconvolved sizes of VLA~1 and VLA~2. 

VLA~1 appears elongated, in the two epochs and at the two wavelengths, in the
SW-NE direction, with a position angle of $\sim 30\arcdeg$, which is in
agreement, within the uncertainties, with the previous measurements reported in
Paper I.  The 2002 epoch 3.6~cm map clearly shows weak emission (at 6~$\sigma$
level) southwestern of VLA~1, \ie\ in the same direction of the elongation of
VLA~1, which forms an apparent one-sided lobe of VLA~1. The 2000 epoch 3.6~cm
map detects this radio lobe only marginally (due to the lower sensitivity).  
The position angle of the VLA~1 one-sided lobe is $\sim 25\arcdeg$, which is
similar to the elongation of VLA~1.  Remarkably, nearly aligned with the major
axis of VLA~1 there are located to the south HH~224NW1 and to the north a chain
of near-IR HH knots (Grosso \et\ 2001).  This coincidence lead Grosso \et\
(2001) to suggest that VLA~1 is powering this HH outflow. 

VLA~2 is also slightly extended, with a deconvolved size of $\sim 0\farcs3$, or
36~AU in projection.  This result is consistent with the upper limit of
$0\farcs4$ obtained in Paper I.  At 6~cm this source is not well separated from
VLA~1,  so the deconvolved size of VLA~2 may not be very accurate.
Fig.~\ref{fmapA} clearly shows that VLA~2 is associated with the near-IR source
and the X-ray emission observed in this region (see \S~\ref{Counter}).   

No radio emission down to 0.05~mJy (3-$\sigma$ level) is detected toward
GY~263, located 6$''$ northwest of VLA~1 and VLA~2.

In order to derive the spectral index of VLA~1 and VLA~2, we first generate
3.6~cm maps using IMAGR and applying a Gaussian taper to the visibility data to
obtain a synthesized beam similar to the 6~cm maps.  Then, the  3.6 and 6~cm
maps were convolved with a Gaussian, so the final resulting beam for the two
wavelengths was the same. For the 2000 epoch, this beam was
$0\farcs90\times0\farcs33, 22\arcdeg$.  For the 2002 epoch, this beam was
$0\farcs67\times0\farcs32, 10\arcdeg$.  Table~\ref{tspi} gives the flux
densities of the convolved maps and their spectral indices for the two epochs. 
The overall radio continuum emission of the YLW~15 region (obtained by
measuring the flux within a box that included  VLA~1, VLA~2 and the weak
extended emission) has an spectral index of 0.5\mm0.1 and does not show signs
of flux variability at roughly a 5\% level.  VLA~1 and VLA~2 have both positive
spectral indices, $\sim$0.5 and $\sim$0.7, respectively. Because both sources
have emission arising from a few tens of AU, their emission is likely optically
partially thick free-free.

\begin{figure}
\plotone{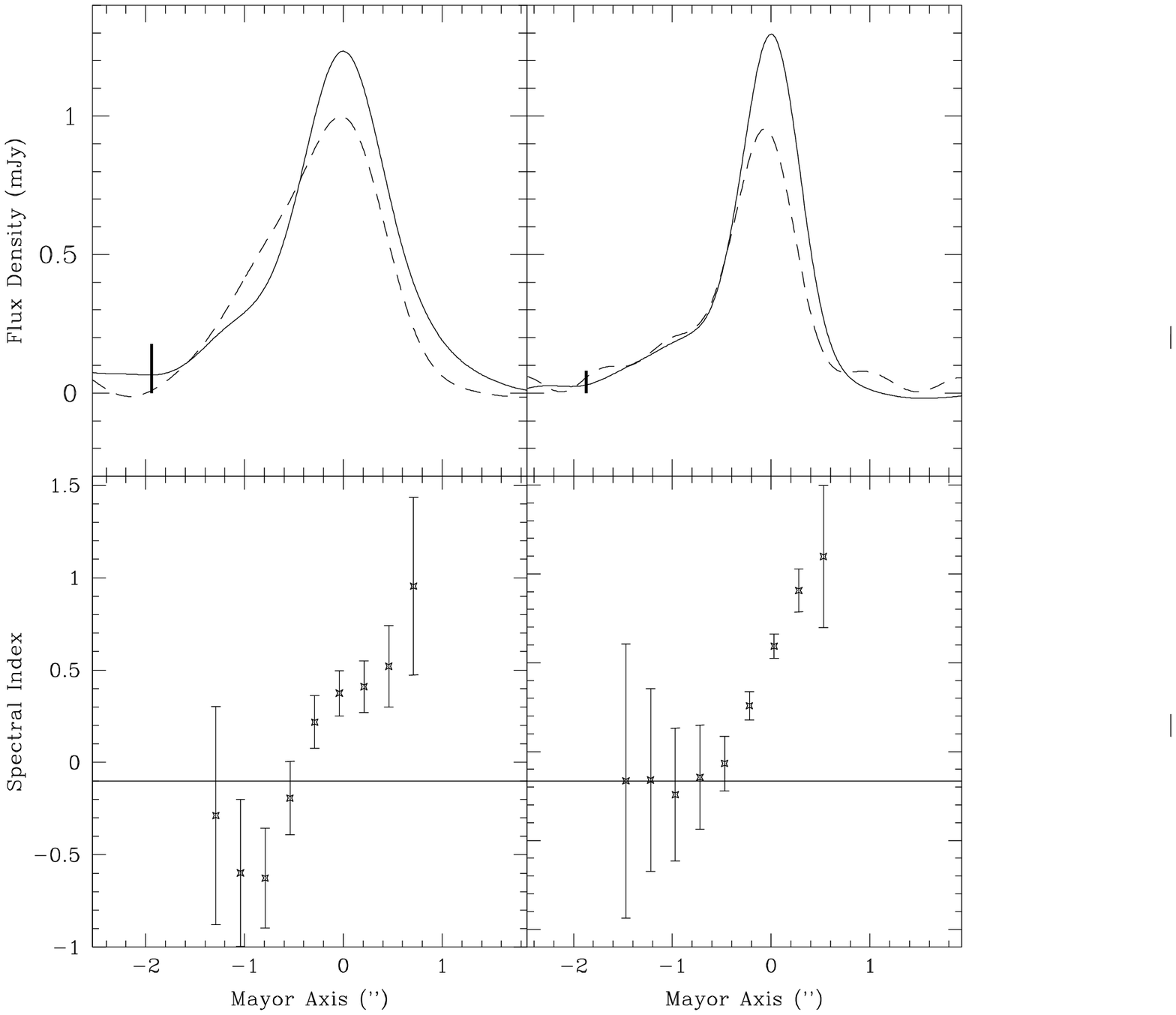}
\caption
{{\em Top panels}: Slices of the VLA~1 emission along its major axis
(PA=25$\arcdeg$) for the epochs 2000 (left panel) and 2002 (right panel).
The solid and dashed lines show the 3.6 and 6~cm emission, respectively.
The small vertical solid line shows the 3-$\sigma$ value for the 6~cm slice.  
{\em Bottom panels}:  Slices of the VLA~1 spectral index between 3.6 and 6~cm
for the epochs 2000 (left panel) and 2002 (right panel). The solid horizontal
line shows the spectral index value of $-$0.1. 
\label{fspi}
}
\end{figure}

Figure~\ref{fspi} shows the slices of the VLA~1 emission and spectral indices
along its major axis for the 2000 and 2002 epoch. Both epochs show that VLA~1
has a spectral index at the peak of $\sim 0.5$.  Interestingly, the 
southwestern weak lobe had in epoch 2000 a spectral index of $-0.6\pm0.3$, 
indicating non-thermal emission, while in the 2002 epoch its spectral index 
raised enough to be compatible with optically thin free-free emission 
($-0.1\pm0.3$).

\subsubsection{VLBA}

No sources were detected from the two epochs of VLBA observations with an upper
limit of about 0.3~\mjy\ (4-$\sigma$ level).  The parallax due to the Earth
orbiting the Sun for $\rho$ Ophiuchi is about 8.3 mas.  YLW~15 has moved 5.4
and -1.4~mas in right ascension and declination, respectively, between the two
VLBA observations due to the parallax.  Taking also into account the spatial
displacement due to the absolute proper motions of VLA~2 measured in Paper II,
the total displacement of VLA~2 is 3.0 and -9.4~mas in right ascension and
declination, respectively.  The visibility data of the first epoch was
corrected for this shift.  The maps of the combined visibility data were
obtained using natural weighting, which yields an rms noise of 52~$\mu$Jy. No
emission was detected down to 0.21~\mjy\ (4-$\sigma$ level). The same process
was done taking into account the parallax and the VLA~1 absolute proper motion.
No emission was detected either. This upper limit implies that at least an 84\%
and 64\% of the radio emission of VLA~1 and VLA~2, respectively, arise from
scales $\ga0.5$~AU, which is in agreement with the VLA results that indicate
that VLA 1 and VLA 2 are being partially resolved at scales of about 20~AU.

\subsubsection{COMICS/SUBARU}

\begin{figure}
\plotone{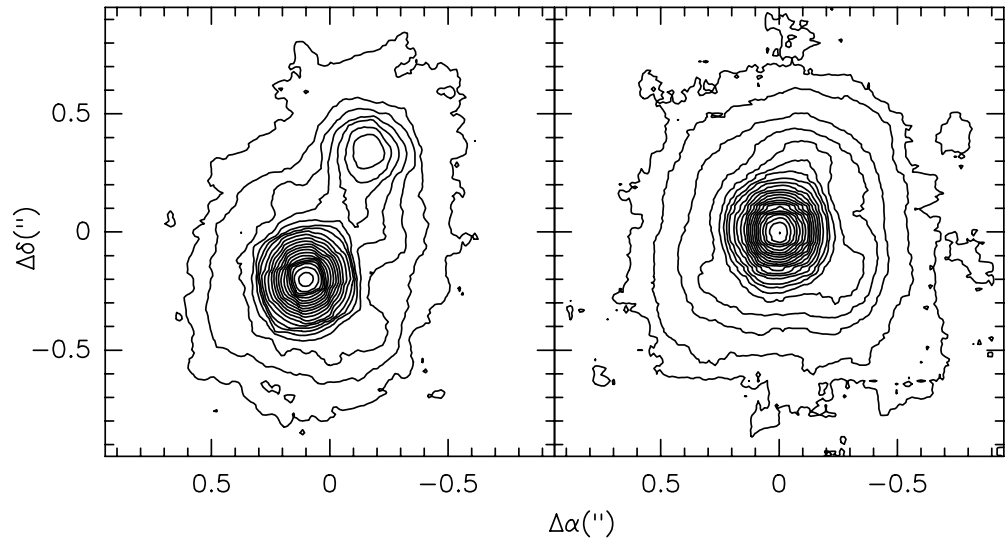}
\caption
{
{\em Left panel}: 
Image of YLW~15 at 8.8~$\mu$m obtained with the COMICS camera of SUBARU.  
{\em Right panel}: 
Image of the standard HD~14~5897 star at 8.8~$\mu$m obtained
with the COMICS camera of SUBARU. Contours in both images are 2, 5, 8, 11,
14, 17, 20 , 25, \nodata, 100 \% of the peak intensity.
\label{fsubaru}
} 
\end{figure}

\begin{figure}
\plotone{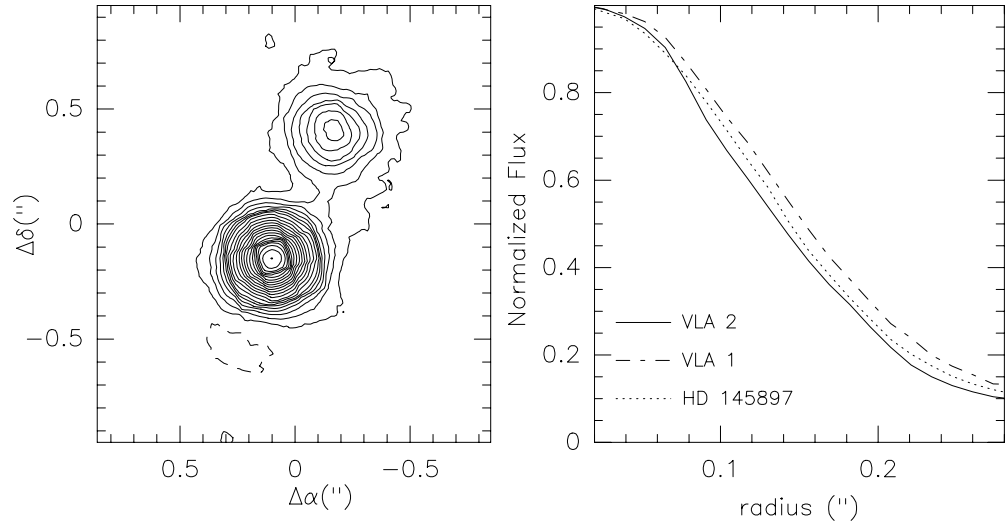}
\caption
{
{\em Left panel}: 
COMICS/SUBARU image of YLW~15, deconvolved (using Maximum Entropy) from the PSF
of the SUBARU telescope at 8.8~$\mu$m, and convolved with a Gaussian beam of
$0\farcs29\times0\farcs27$ and PA=$-2\arcdeg$.  Contours are 2, 5, 8, 11,
14, 17, 20 , 25, \nodata, 100 \% of the peak intensity of VLA~2.  
{\em Right panel}: 
Annular radial averaged profile of HD~14~5897 (dotted line), of YLW~15 VLA~2
(solid line) and VLA~1 (dashed line) of the normalized flux. 
\label{fdcnv}
} 
\end{figure}

The left and right panels of Figure~\ref{fsubaru} show the image of YLW~15 and
the standard HD~14~5897, respectively, at 8.8~$\mu$m obtained with the COMICS
camera of SUBARU.  The two sources in YLW~15 are well separated in the image,
with VLA~2 being significantly stronger than VLA~1.  The separation between
VLA~1 and VLA~2 is about $0\farcs59$ with a PA of $\sim 147\arcdeg$, which is
in agreement with the values obtained from the VLA observations (see Table 3
from Paper II).  There is extended emission observed up to a $\sim 8$\% level
with respect to the peak value of the image around the two sources. It is clear
from the HD~14~5897 image that there is also an  extended ''halo'' at a level
of 14\% level due to the PSF of the SUBARU telescope.  Therefore the extended
emission up to a $\sim 8$\% level in the YLW~15 image is possibly due to the
SUBARU PSF. In order to confirm this possibility, we deconvolved the YLW~15
image using as a PSF model the HD~14~5897 image masked for levels below 2\% (in
order to avoid the noise to contribute into the deconvolution). Three different
procedures were used for the deconvolution: the Lucy algorithm from the STSDAS
package of IRAF, Maximum Entropy and CLEAN, these two from the MIRIAD package. 
The left panel of Figure~\ref{fdcnv} shows the resulting image of the Maximum
Entropy deconvolution: the extended emission has disappeared down to a 2\%
level (a similar result is also obtained with the Lucy and CLEAN algorithms).
Thus, it is clear that the extended halo around YLW~15 observed in the original
image was due to the SUBARU PSF. 

In order to check if VLA 1 and VLA 2 are resolved at 8.8~$\mu$m we computed the
radial profile of the annular averaged normalized flux of HD~14~5897, VLA~2 and
VLA~1 (see the right panel of Figure~\ref{fdcnv}).  The half-width half-maximum
(HWHM) of these profiles are $0\farcs143$, $0\farcs138$ and $0\farcs154$ for
HD~14~5897, VLA~2 and VLA~1, respectively. HD~14~5897 should be unresolved (its
size is $\sim2$~mas: Cohen \et\ 1999). The slightly higher width of
HD~14~5897 with respect to VLA~2 could be due to a slightly higher seeing
during the HD~14~5897 observations (taken 30 minutes after YLW~15).  In any
case, the profile of VLA~2 likely suggests that it is unresolved ($\theta_{\rm
FWHM}({\rm VLA~2}) \ll 0\farcs28$).  Assuming, that VLA~2 is unresolved, and
that it traces the PSF, then the deconvolved HWHM of VLA~1 can be estimated
from the observed HWHM of these two sources: $\sqrt{0.154^2 - 0.138^2} \simeq
0.068$.  There are two sources of uncertainties in this estimation. First, the 
intrinsic uncertainty due to the angular resolution and the SNR of the source: 
$FWHM/SNR$ (Condon 1997). Since the SNR for VLA~1 is $\simeq 20$ and the
angular resolution  is $FWHM\simeq0\farcs28$, the uncertainty is $0\farcs014$. 
Second, the difference in the measured size of the reference star and VLA 2
can also be accounted as source of uncertainty: $\simeq0\farcs005$. The total
uncertainty  is $\sqrt{0.014^2+0.005^2}\simeq0\farcs015$.  Thus, the FWHM size
for VLA~1 of $0\farcs14\pm0\farcs03$ or $16\pm4$~AU.  Finally, we note that
despite the aforementioned deconvolution algorithms are not being very reliable
to derive the YLW~15 sizes because of the small difference in size between
VLA~2 and HD~14~5897, all of them produce an unresolved deconvolved object for
VLA~2 (assuming that HD~14~5897 is a point source),  whereas for VLA~1 they
produce a resolved object  with a FWHM size of $0\farcs10$--$0\farcs14$, in
agreement with the previous radial profile analysis.

The slightly resolved VLA~1 mid-IR emission possibly traces matter in a
non-disk 3-D distribution (\ie\ from the envelope) which is falling onto (or
perhaps is being flung off of) its circumstellar disk (e.g. Chick, Pollack, \&
Cassen 1996; Adams, Lada, \& Shu 1987).   Yet, for low inclination angles
(nearly face-on) the circumstellar disk emission could also contribute to the
mid-IR emission (Osorio \et\ 2003; Osorio: Private communication).  Indeed, CO
outflow observations suggest that YLW~15 is nearly face-on (Bontemps \et\
1996). If this mid-IR emission comes partially from a circumstellar disk, then,
it will trace only the inner-region of the disk which is warm enough to radiate
in these wavelengths.  Since the infalling envelope also contributes to the 
mid-IR emission, the size of the mid-IR emission does not correspond with the 
disk size directly.  However, due to tidal truncation from the companion star 
(located at 72~AU in projection) its true size is probably not much larger 
(\eg\ \ro\ \et\ 1998; Loinard \et\ 2002).  On the other hand, the mid-IR 
emission of VLA~2, unresolved, arises from a significantly smaller 
circumstellar structure.

\subsubsection{Counterparts of the VLA sources\label{Counter}}

YLW~15 is detected in the 2MASS All Sky Survey in the J and K bands, and is
cataloged as 2MASS~1627269-244050. The 2MASS (second incremental release) Point
Source Catalog coordinates are given in Table~\ref{tcomp}. These near-IR data
were obtained on 1998 July 9 (1998.52). In order to properly compare the
positions of 2MASS~1627269-244050 with VLA~1 and VLA~2, we corrected the
positions of these two radio sources to the values expected for the 1998.52
epoch taking into account the proper motions measured in Paper II. The
difference between the positions of VLA~1 and VLA~2 for the 1998.52 epoch with
respect to the 2MASS position are given in Table~\ref{tcomp}.  From this table
it is clear that the near-IR source is associated with VLA~2, as suggested by
Greene \& Lada (2002). Similarly, we compared the positions of the two VLA
sources from the 2000 observations with the position of the X-ray source
obtained with the CHANDRA satellite in 2000 (Imanishi, Koyama \& Tsuboi 2001;
Imanishi, private communication).  Given the positions uncertainties, the X-ray
emission is more likely associated with VLA~2.

High angular resolution, near-IR, observations with HST/NICMOS (Allen \et\
2002) and using lunar occultation techniques (Simon \et\ 1995) show that YLW~15
has a companion 6$''$ to the NW (720~AU): GY~263.  Yet YLW~15 appears to be a
single object at scales down to $0\farcs1$.  Therefore, given the clear
association of VLA~2 with the 2MASS~1627269-244050 source, VLA~2 is the only
significant source of emission at near-IR wavelengths.  Based on the near-IR
spectroscopy, VLA~2 is a K5 IV/V YSO of 0.5~\mo\ (Greene \& Lada 2002).   The
mid-IR images at 8.8~$\mu$m from COMICS with SUBARU (this paper) and at
10.8~$\mu$m (Haisch \et\ 2002) show two sources that correspond to VLA~1 and
VLA~2.  At these two wavelengths, VLA~2 is stronger than VLA~1, which suggests
that VLA~1 is an embedded object or, alternatively, is less luminous 
(see \S~\ref{edat}).

\subsection{YLW~16A}

\subsubsection{VLA}

\begin{figure}
\plotone{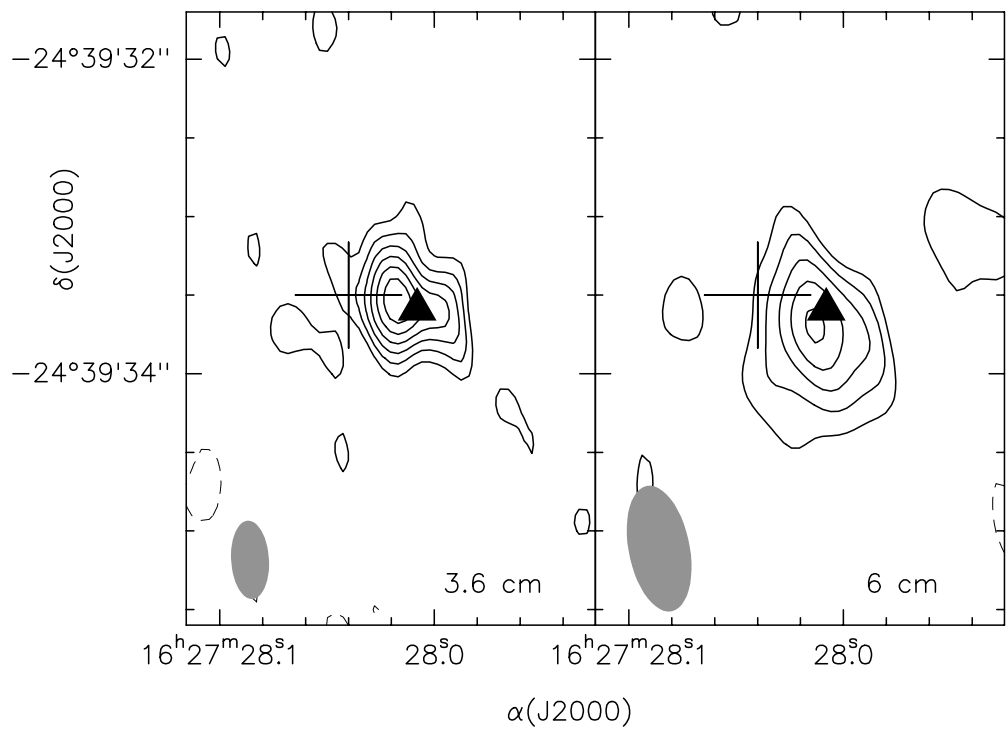}
\caption
{VLA continuum images of YLW~16A at 3.6~cm (left panel) and 6~cm (right panel)
for the epoch 2002.  Contours are $-$2, $-$3, 2, 3, 4, 5, 6, 7~$\times$ the
rms noise of the image (17 and 21~\Mjy at 3.6 and 6~cm, respectively). The
half power size of the synthesized beam ($0\farcs40\times0\farcs19$,
PA=2$\arcdeg$ and $0\farcs81\times0\farcs31$, PA=12$\arcdeg$ at 3.6 and 6~cm,
respectively) is shown in the bottom left corner. The cross shows the position
uncertainty of the X-ray source no. 57 (Imanishi et al. 2001).  The filled
triangle shows the position of the near-IR 2MASS~1627280-243933 source.
\label{fmapB}
}  
\end{figure}

This source is located 78$''$ north of YLW~15, so it is far from the phase
center of the VLA observations, but still within the primary beam of the 3.6
and 6~cm maps.  This source was only detected in the 2002 epoch. Because this
source is weak, in order to achieve a better SNR, the maps presented here were
obtained using robust weighting of 0 and 1 at 3.6 and 6~cm, respectively.
Figure~\ref{fmapB} shows the 3.6~cm and 6~cm maps for YLW~16A. The 3.6~cm image
shows that the emission is elongated to the west of the intensity peak. Because
of the low SNR, it is not clear whether this elongation is due to a one side
jet lobe (as it happens in YLW~15 VLA~1) or if it is due to a binary radio
source separated by $\sim0\farcs3$.

Because bandwidth smearing is important for YLW~16A at 3.6~cm ($\sim 0\farcs5$
in the north-south direction),  only the integrated flux density was measured,
using a box that included all the emission.  Thus, the primary beam corrected
flux density of YLW~16A is 0.43\mm0.04 and 0.27\mm0.05~mJy at 3.6 and 6~cm,
respectively.  These values imply an  spectral index of the overall emission of
0.8\mm0.4, suggesting a thermal radio jet origin.

\subsubsection{Counterparts of the VLA sources}

Allen \et\ (2002) detected with the HST/NICMOS3 camera two near-IR sources 
with their components separated by $\sim 0\farcs5$ in the east-west direction. 
However, Allen \et\ (2002) suggested that this apparent binary system is
possibly tracing scattered light from the disk of a single YSO. In addition,
lunar occultation observations by Simon \et\ (1995) failed to detect a binary
system.  To compare the position of the near-IR and VLA emission, we used the
position given by the 2MASS Point Source Catalog (source 2MASS
1627280-243933).  Correcting the VLA position for proper motions (as we did in
\S~\ref{Counter}), the difference between the VLA and near-IR emission, $\Delta
\alpha = 0\farcs10\pm0\farcs19$,  $\Delta \delta = 0\farcs17\pm0\farcs17$, 
implies that both wavelength trace the same source. The CHANDRA X-ray source n.
57 (Imanishi, private communication) also coincides within its position
uncertainty ($\sim 0\farcs5$) with the VLA peak emission.

\section{Discussion\label{disc}}

\subsection{The Radio Continuum properties}

The jet-like morphology of VLA~1 (including its weak southwestern lobe) and its
spectral index at centimeter wavelengths, $\sim 0.5$, suggests that this source
is a thermal radio jet (\eg\ Anglada 1996; \ro\ 1997).  VLA~2 has a slightly
higher spectral index, $\sim 0.7$, and it is also partially resolved at 
similar scales as VLA~1. Yet, it does not have a clear elongation direction
(see Table~\ref{tflux}).  Since VLA~1 and VLA~2 are very young stellar objects
(see \S~\ref{edat}) and that a significant fraction of their radio emission
arises  at scales of $\ga 1$~AU (\S~\ref{vla}), we can assume that their 3.6
and 6~cm  emission arises from collimated, ionized winds. The radio emission of
these winds was modeled by Reynolds (1986), which predicts that for a wind with
constant velocity, temperature and ionization, a lower spectral index implies
that the collimation of the jet increases with distance to the exciting star. 
Thus the radio continuum properties of the YLW~15 radio sources (VLA~1 has a
lower spectral index and has a more clear jet-like morphology than VLA~2) are
in agreement with the expected properties of collimated ionized winds. The
higher 3.6~cm flux of VLA~1 with respect to VLA~2 implies that the momentum
rate of the outflow material is higher in VLA~1 than in VLA~2.  From the
correlation between the molecular outflow momentum rate and the flux of the
free-free emission estimated by Anglada \et\ (1998), the momentum rate expected
for the molecular outflow powered by VLA~1 and VLA~2 should be 5\xx10$^{-5}$
and 2\xx10$^{-5}$~\mo~yr$^{-1}$~\kms, respectively.  On the other hand, the
lack of emission for GY~263 gives an upper limit for the momentum rate of
1\xx10$^{-6}$~\mo~yr$^{-1}$~\kms, more than an order of magnitude lower than
that of VLA~2. This suggests that GY~263 is a more evolved YSO.  Indeed, the
near-IR to mid-IR spectral index of GY~263 is consistent with a Class II YSO 
(Haisch et al. 2002).

\subsection{The Radio Continuum and X-ray Emissions}

VLBA observations do not detect any compact emission in YLW~15 down to $\sim
0.21$~\mjy.  Since VLA~2 is associated with X-ray emission, it is expected to
have milliarcsecond radio emission.   Yet, the partially extended subarcsecond
free-free emission associated with VLA~2 will significantly mask emission at
milliarcsecond scales:  the observed angular sizes at 6~cm suggest that the
radio ``photosphere'' has dimensions $\geq$10 AU.  Any emission processes
taking place at 6~cm inside this photosphere will suffer from significant
free-free absorption and will be very difficult to detect. In order to avoid
the absorption of the extended free-free emission, and thus to be more
sensitive to milliarcsecond emission, VLBA observations would be best done at
2~cm (where the radio photosphere will be about two times smaller and the
optical depth will also be smaller) or at shorter wavelengths.

\subsection{The evolutionary status of VLA~1 and VLA~2\label{edat}}

\begin{figure}
\plotone{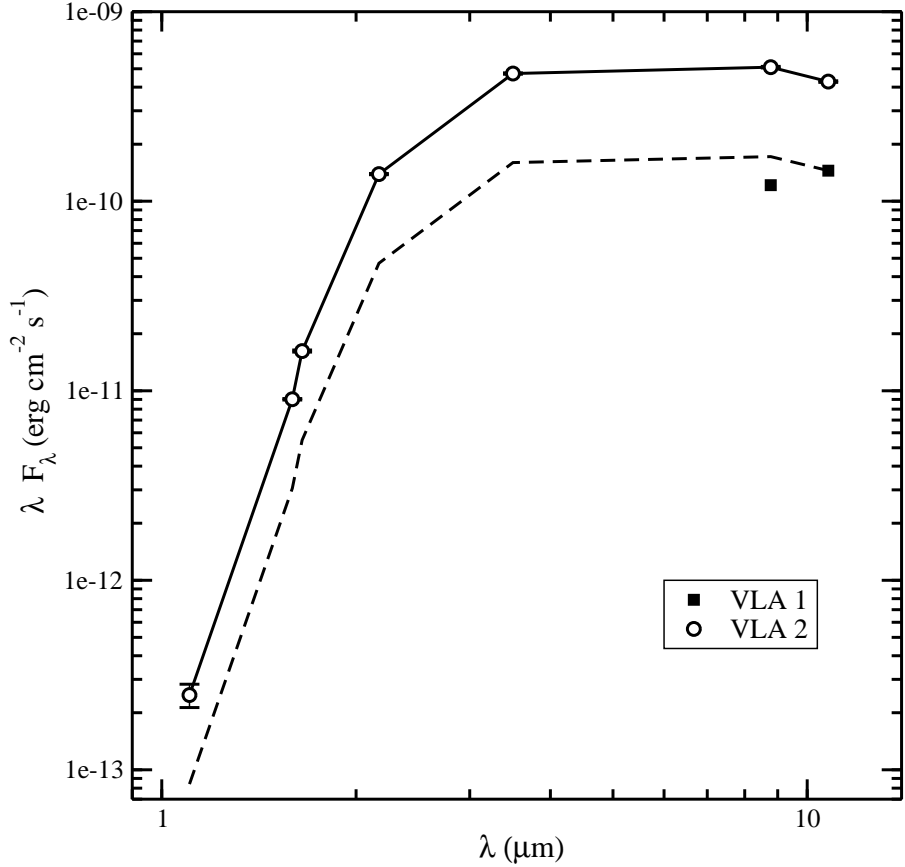}
\caption
{Spectral energy distribution (SED) in the near through mid IR of the YLW~15
VLA~1 and VLA~2 emission. Data at 1.1 and 1.6~$\mu$m from Allen et al. (2002).
Data at 1.65, 2.17, 3.5 and 10.78~$\mu$m from Haisch et al. (2002). Data at
8.8~$\mu$m from the SUBARU image presented in this paper. The solid line
follows the SED of VLA~2. The dashed line is the VLA~2 SED, scaled down by a
factor 2.95, so the $\lambda F_{\lambda}$ value of the dashed line at
10.78~$\mu$m coincides with that of VLA~1.  
\label{fsed}
}  
\end{figure}

Greene \& Lada (2002) and Haisch et al. (2002) argue that VLA~1 and its
mid-IR counterpart is not a true star but instead has an outflow origin
(\ie\ it could be an HH knot).  However, the VLA proper motions measured
in VLA~1 and VLA~2 (Paper II) clearly shows that these two sources are
gravitationally bounded, with VLA 1 being more massive. Therefore, it is
clear that VLA~1 is a YSO. The properties of VLA~1 and VLA~2 can be
derived from the near-IR through far-IR properties of YLW~15 available in
the literature.  Figure~\ref{fsed} shows the $\lambda F_{\lambda}$ --
$\lambda$ plot for those observations in the near to mid-IR range with
enough angular resolution to resolve spatially VLA~1 and VLA~2 (Allen \et\
2002; Haisch \et\ 2002).  It is clear from this plot that the luminosity
in the near to mid-IR is clearly dominated by VLA~2. The dashed line shows
the spectral energy distribution (SED) of VLA~2 scaled down, so the
10.8~$\mu$m flux coincides with the VLA~1 flux at this wavelength. The 1.6
(Allen \et\ 2002) and 2.2~$\mu$m (Haisch \et\ 2002) observations have
enough angular resolution to resolve the binary system, but only VLA~2 is
detected. The presence of significant scattered light around YLW~15 does
not allow to obtain useful upper limits for VLA~1 (the 2.2~$\mu$m upper
limit for VLA~1 is only 0.2~mag higher than the VLA~2 magnitude, Haisch,
private communication). In other words, all what we can say is that in the
near to mid-IR wavelengths VLA~1 is less luminous than VLA~2. 

The observed near-IR properties show that VLA~2 is a Class I YSO (Haisch
\et\ 2002;  Greene \& Lada 2002).  In addition, the overall properties of
YLW~15 from the far-IR through the mm wavelengths (see \S~\ref{intro})
show they also are in agreement with those found in Class I objects.  Yet,
which is the evolutionary state of VLA~1? Taking into account the
near-to-mid IR SED (Fig.~\ref{fsed}), there are two possible scenarios for
VLA~1: it is more embedded than VLA~2 (i.e., with a steeper SED), or,
alternatively, VLA~1 has a similar SED but a lower bolometric luminosity
than VLA~2. For this later scenario, the upper limit of the VLA~1 stellar
luminosity would be 3~\lo\ (the value for VLA~2: Greene \& Lada 2002).  In
the first scenario, since the overall properties of YLW~15 are in
agreement with those of Class I sources, VLA~1 cannot be much more
embedded than VLA~2 (\ie\ a Class 0 object). We suggest that VLA~1 is a
very embedded YSO in the Class 0 to Class I transition, as it may also be
L1448~IRS3(A) (Ciardi \et\ 2003).  In this scenario, VLA~1 probably
dominates the millimeter dust emission and has a stronger outflow. 
Indeed, VLA~1 is the powering source of a HH system (Grosso \et\ 2001) and
its radio centimeter emission is stronger than in VLA~2.  We speculate
that being YLW~15 a binary young stellar system, the dynamical
perturbation of the two YSO could diminish the circumstellar mass with
respect to the expected values for isolated stars. In fact, the more
evolved, Class II, T Tauri close binary stars (with separations in the
1-100~AU range) have less millimeter emission than single T Tauri stars
(Jensen, Mathieu \& Fuller 1996). 

A conclusive test of which of the two sources is more luminous and which is 
the evolutionary status of VLA~1 as well as its dust properties will come 
from high angular resolution observations in the mm and sub-mm, as those that
will be achieved in the future by the SMA, CARMA and ALMA, and at other 
wavelengths in the near to mid-IR range.

The most striking feature of the YLW~15 binary system is that VLA~1
appears to be a more massive and more embedded (thereby less evolved) or,
alternatively, less luminous YSO than VLA~2.  Interestingly, this result
is apparently against the expected evolutionary scenario, where one
expects that the more massive YSO in a binary system is the more evolved
and luminous YSO.  Simulations of the evolution of binary YSOs with
circumstellar and circumbinary disks show that if one of the two YSO is
less massive, the dynamical effects of the system can cause a larger
accretion rate to the less massive star (that is expected to evolve more
slowly), which will tend to equalize the masses (Lubow \& Artymowicz
2000).  But, this still does not explain that the more massive is
apparently less evolved or less luminous.  A plausible explanation is 
that the less massive component may indeed be more luminous because of its 
higher accretion rate, since (accretion) luminosity is proportional to accretion 
rate times stellar mass. GY~263, a Class II YSO and so
clearly more evolved than the YLW~15 binary system, is only at 6$''$
(720~AU in projection) from YLW~15.  If it is truly so close to YLW~15,
GY~263 could have had in the past a close approach to YLW~15 and altered
significantly the star formation process of YLW~15, possibly reducing the
accretion process to the circumbinary disk around YLW~15 (which may
explain the low millimeter flux of YLW~15 and the low momentum flux of the
CO outflow) and, therefore, altering somewhat the evolutionary scenario of
the binary system.

\section{Conclusions}

We have carried out VLA and VLBA radio as well as COMICS/SUBARU mid-IR
(8.8~$\mu$m)  observations of the YSO binary system in YLW~15 and carried out a
study of the properties of YLW~15 from previous results available in the
literature. The main conclusions are:

\begin{enumerate}
\item The properties of the radio continuum emission of VLA~1 and VLA~2 show
that these two sources have partially thick free-free emission, likely due to
collimated, ionized winds, as it is the case of other Class I and 0 YSOs, with
associated molecular and/or HH outflows.
\item The centimeter VLA~1 emission has a jet-like morphology, with the axis
coinciding in position angle with the HH outflow found by Grosso et al. (2001)
and a jet length of $\sim 0\farcs32$ or $\sim 40$~AU. Undetected in the
near-IR, becomes ''visible'' at 8.8~$\mu$m. The 8.8~$\mu$m emission is slightly
resolved, with a deconvolved size of $\sim 16$~AU, suggesting that the emission
arises from a compact circumstellar structure around the protostar.
\item The centimeter VLA~2 emission is partially resolved at scales of 20~AU.
Comparison with data taken at other wavelengths shows that VLA~2 is the
responsible of the near-IR emission and it is the source of the strong X-ray
emission.  This source appears unresolved at 8.8~$\mu$m, implying a small
circumstellar structure.
\item VLBA observations failed to detect continuum emission at 6~cm. The lack
of ''peri-stellar'' emission is likely due to a high optical depth of the
free-free emission, that arises at larger scales. Shorter wavelengths should be
used to detect the expected centimeter emission associated with the X-ray
emission.
\item The near to mid-IR properties of YLW~15 show that VLA~1 is less luminous
than VLA~2 in this wavelength range. This could be due to a lower bolumetric
luminosity of VLA~1 compared with VLA~2 or alternatively indicate that VLA~1 
is a more embedded YSO, and thus, younger or less evolved, than VLA~2. 
VLA~2 has the SED properties of a typical Class I YSO. Therefore, VLA~1 could 
be either in the same evolutionary state than VLA~2 or either in the Class 0 
to Class I transition. 
\item VLA~1 appears to be more massive but more embedded or,
alternatively, less luminous than VLA~2. This result is apparently against
the expected evolutionary scenario, where one expects that the more
massive YSO in a binary system is also the more evolved and luminous YSO. 
\item YLW~16A is detected with the VLA. Its position coincides well with the
near-IR and X-ray emission observed in the region.
\end{enumerate}

\acknowledgments

We would like to thank the anonymous referee for the valuable comments.
This publication makes use of data products from the Two Micron All Sky Survey,
which is a joint project of the University of Massachusetts and the Infrared
Processing and Analysis Center/California Institute of Technology, funded by
the National Aeronautics and Space Administration and the National Science
Foundation.  JMG acknowledges support by RED-2000 from the Generalitat de
Catalunya, by DGICyT grant AYA2002-00205 and by the Divisi\'o de Ci\`encies
Experimentals i Matem\`atiques of the Univ. of Barcelona. JMG thanks the
hospitality and support of the Instituto de Astronom\'{\i}a-UNAM. SC
acknowledges support from CONACyT grant 33933-E and JC from CONACyT grants
34566-E and 36572-E.  LFR acknowledges support from CONACyT, M\'exico and
DGAPA, UNAM.



\clearpage
\begin{deluxetable}{ccccc}
\tabletypesize{\scriptsize}
\tablecaption{VLA Observations of YLW 15\label{tobsA}}
\tablewidth{0pt}
\tablehead{
\multicolumn{1}{c}{} &
\multicolumn{2}{c}{Synthesized} &
\multicolumn{2}{c}{rms Noise} 
\\
\multicolumn{1}{c}{} &
\multicolumn{2}{c}{Beam}  &
\multicolumn{2}{c}{(\Mjy)}   
\\
\multicolumn{1}{c}{Epoch} &
\multicolumn{2}{c}{\hrulefill} &
\multicolumn{2}{c}{\hrulefill} 
\\
\multicolumn{1}{c}{(yr)} &
\multicolumn{1}{c}{3.6 cm} &
\multicolumn{1}{c}{6 cm} &
\multicolumn{1}{c}{3.6 cm} & 
\multicolumn{1}{c}{6 cm}
}
\startdata
(2000)&0\farcs55\xx0\farcs19,+27&0\farcs90\xx0\farcs33,+25&50&59 \\
(2002)&0\farcs40\xx0\farcs19,+2 &0\farcs67\xx0\farcs31,+10&17&26 \\
\enddata
\end{deluxetable}

\begin{deluxetable}{cccc}
\tabletypesize{\scriptsize}
\tablecaption{VLBA Observations of YLW 15\label{tobsB}}
\tablewidth{0pt}
\tablehead{
\multicolumn{1}{c}{Epoch} &
\multicolumn{1}{c}{Phase} &
\multicolumn{1}{c}{rms Noise} & 
\multicolumn{1}{c}{Synthesized}
\\
\multicolumn{1}{c}{(yr)} &
\multicolumn{1}{c}{Calibrator} &
\multicolumn{1}{c}{(\Mjy)} & 
\multicolumn{1}{c}{Beam (mas)}
}
\startdata
(2001)&J1625-2527&71&6.1\xx2.2,-6 \\
(2002)&J1625-2527&74&4.9\xx1.9,-2  \\
\enddata
\end{deluxetable}

\begin{deluxetable}{cccc@{\hspace{0.2cm}}c@{\hspace{0.0cm}}ccc@{\hspace{0.2cm}}c@{\hspace{0.0cm}}c}
\tabletypesize{\scriptsize}
\tablecaption{VLA Results\label{tflux}}
\tablewidth{0pt}
\tablehead{
\multicolumn{1}{c}{} &
\multicolumn{1}{c}{$\alpha$(J2000)} &
\multicolumn{1}{c}{$\delta$(J2000)} &
\multicolumn{3}{c}{3.6 cm} &
\multicolumn{1}{c}{} &
\multicolumn{3}{c}{6 cm}  
\\
\cline{4-6}
\cline{8-10}
\multicolumn{1}{c}{Year} &
\multicolumn{1}{c}{($16^{\rm h}27^{\rm m}$)} &
\multicolumn{1}{c}{($-24\arcdeg40'$)} &
\multicolumn{1}{c}{S$_{\nu}$ (mJy)} &
\multicolumn{1}{c}{Deconvolved Size} &
\multicolumn{1}{c}{PA} &
\multicolumn{1}{c}{} &
\multicolumn{1}{c}{S$_{\nu}$ (mJy)} &
\multicolumn{1}{c}{Deconvolved Size} & 
\multicolumn{1}{c}{PA} 
}
\startdata
\multicolumn{2}{l}{YLW~15 VLA 1:} \\
2000 & 26.9120\mm0.0004 & 50.295\mm0.013 &
      1.40\mm0.10 & \DS{0}{40}{4} \xx \DS{0}{07}{2}  &  \D{30}{4} &&
      1.41\mm0.13 & \DS{0}{79}{9} \xx \DS{0}{13}{5} &  \D{26}{3}  \\
2002 & 26.9125\mm0.0002 & 50.331\mm0.005 &
      1.51\mm0.04 & \DS{0}{32}{1} \xx $\la0\farcs08$& \D{35}{3} &&
      1.38\mm0.06 & \DS{0}{51}{3} \xx \DS{0}{11}{5} & \D{33}{3} \\
\multicolumn{2}{l}{YLW~15 VLA 2:} \\
2000 & 26.9342\mm0.0008 & 50.795\mm0.027 & 
      0.78\mm0.11 & \DS{0}{35}{8}  \xx \DS{0}{12}{4} & \D{26}{12} &&
      0.74\mm0.13 & \DS{0}{56}{15} \xx $\la0\farcs21$& \D{65}{15} \\
2002 & 26.9294\mm0.0004 & 50.899\mm0.012 & 
      0.64\mm0.04 & \DS{0}{26}{3} \xx \DS{0}{20}{4} & \D{11}{30} &&
      0.55\mm0.05 & \DS{0}{27}{4} \xx $\la0\farcs16$& \D{4}{21} \\
\multicolumn{2}{l}{YLW~16A:} \\
2002 & 28.013\mm0.005 & 33.69\mm0.07 & 
      0.55\mm0.05 & \nodata\tablenotemark{a} & \nodata &&
      0.27\mm0.05 & \nodata\tablenotemark{a} & \nodata \\
\enddata
\tablenotetext{a}{Possibly a binary source. See discussion in text.}
\end{deluxetable}

\begin{deluxetable}{cccccccc}
\tabletypesize{\scriptsize}
\tablecaption{YLW 15: Flux Densities and Spectral 
Indices\tablenotemark{a}\label{tspi}}
\tablewidth{0pt}
\tablehead{
\multicolumn{1}{c}{} &
\multicolumn{3}{c}{2000} &
\multicolumn{1}{c}{} &
\multicolumn{3}{c}{2002}
\\
\cline{2-4}
\cline{6-8}
\multicolumn{1}{c}{Source} &
\multicolumn{1}{c}{S$_{3.6 cm}$} &
\multicolumn{1}{c}{S$_{6 cm}$} & 
\multicolumn{1}{c}{$\alpha$} &
\multicolumn{1}{c}{} &
\multicolumn{1}{c}{S$_{3.6 cm}$} & 
\multicolumn{1}{c}{S$_{6 cm}$} &
\multicolumn{1}{c}{$\alpha$} 
}
\startdata
VLA 1 & 1.88\mm0.10&1.38\mm0.12&0.56\mm0.18 && 
        1.66\mm0.03&1.27\mm0.05&0.48\mm0.08 \\
VLA 2 & 0.88\mm0.10&0.54\mm0.11&0.88\mm0.41 &&
        0.84\mm0.04&0.57\mm0.05&0.70\mm0.18 \\
Whole Region & 
        2.96\mm0.17&2.30\mm0.22&0.46\mm0.20 &&         
        2.85\mm0.08&2.18\mm0.12&0.48\mm0.11 \\ 
\enddata
\tablenotetext{a}{Flux densities in mJy}
\end{deluxetable}

\begin{deluxetable}{cccccc}
\tabletypesize{\scriptsize}
\tablecaption{2MASS and CHANDRA positions\label{tcomp}}
\tablewidth{0pt}
\tablehead{
\multicolumn{2}{c}{} &
\multicolumn{1}{c}{$\alpha$(J2000)} &
\multicolumn{1}{c}{$\delta$(J2000)} &
\multicolumn{2}{c}{Offset Positions ($\Delta \alpha$,$\Delta \delta$)} 
\\
\multicolumn{1}{c}{Object} &
\multicolumn{1}{c}{Epoch} &
\multicolumn{1}{c}{16$^{\rm h}$27$^{\rm m}$} &
\multicolumn{1}{c}{$-24\arcdeg$40$'$} &
\multicolumn{1}{c}{{\em wrt} VLA~1\tablenotemark{a}} &
\multicolumn{1}{c}{{\em wrt} VLA~2\tablenotemark{a}}
}
\startdata
2MASS~1627269-244050&1998.52& 26.944\mm0.012 & 50.73\mm0.13 &
     0.42\mm0.17,0.48\mm0.14& 0.18\mm0.17,0.06\mm0.16 \\
CHANDRA no. 54      &2000.28& 26.950\mm0.037 & 50.9\mm0.5   & 
        0.5\mm0.5,0.6\mm0.5 & 0.2\mm0.5,0.1\mm0.5 \\
\enddata
\tablenotetext{a}{{\em wrt}: with respect to}
\end{deluxetable}

\end{document}